\documentclass[A4]{spie}  
\usepackage[]{graphicx}

\title{Hidden photon measurements using the long-baseline cavity of laser interferometric gravitational-wave detector
}

\author{Y. Inoue\supit{a}, K. Ishidoshiro\supit{b}
\skiplinehalf
\supit{a}The Graduate University for Advanced Studies, Kanagawa, Japan\\
\supit{b} Tohoku univercity, Miyagi, Japan
}

\authorinfo{Further author information: (Send correspondence to Yuki Inoue)\\Yuki Inoue: E-mail: iyuki@post.kek.jp, Telephone: +81-29-864-5200 ex 2711}

\begin{document} 
  \maketitle 

\begin{abstract}
We suggest a new application for the long-baseline and high powered cavities in a laser-interferometric gravitational-wave~(GW) detector to search for WISPs (weakly interacting sub-eV particles), such as a hidden U(1) gauge boson, called the hidden-sector photon. 
It is based on the principle of a light shining through the wall experiment, adapted to the laser with a wavelength of 1064 or 532~nm.
The transition edge sensor (TES) bolometer is assumed as a detector, which the dark rate and efficiency are assumed as $0.000001~\mathrm{s^{-1}}$ and 0.75, respectively.
The TES bolometer is sufficiently sensitive to search for the low-mass hidden-sector photons.
We assume that the reconversion cavity is mounted on the reconversion region of hidden-sector photons, which number of reflection and length are assumed as 1000 and 10, 100, and 1000m.
We found that the second-point-five and the second generation GW experiments, such as KAGRA and Advanced LIGO with a regeneration cavity and TES bolometers.
The expected lower bounds with these experiments wit the reconverted mirror are set on the coupling constant $\chi = 2 \times 10^{-9}$ for hidden-sector photon with a mass of $2 \times 10^{-5}$~eV within 95\% confidence level. The third generation detector, Einstein Telescope, will reach $\chi = 1 \times 10^{-9}$ at a mass of $1 \times 10^{-5}$~eV within 95\% confidence level. 
Although the operation and construction of the RC will demand dedicated optical configurations, the cavities used in GW detection are expected to measure the strong potential for finding the hidden-sector photons.
\end{abstract}

\section{Introduction}  
\begin{figure}
\begin{center}
\includegraphics[width=12cm]{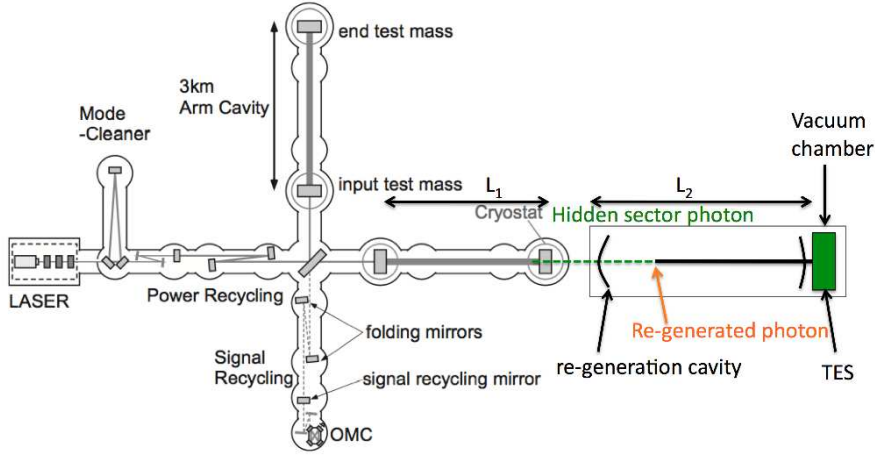}
\caption{Schematic of the LSW gravitational wave experiment. The vacuum chamber is placed at the back of arm of the cavity. The RC is mounted in the vacuum chamber (from KAGRA~\cite{Somiya:2011np,Aso:2013eba}).  \label{fig:LSW} }
\end{center}
\end{figure}

The theories of beyond the standard-models predict an extra U(1) gauge corresponding to the hidden sectors~\cite{Ahlers:2007rd,PhysRevLett.38.1440, PhysRevLett.40.223,Kim19871,PhysRevLett.40.279}. Low-mass particles residing in the hidden sectors are expected to weakly interact with the visible sectors; hence they are  called WISPs~(weakly interacting sub-eV particles).
In an experimental search, WISPs should be revealed by weak kinetic mixing between a photon $\gamma$ and a hidden-sector photon $\gamma'$~\cite{Dienes:1996zr}.
The probability of a $\gamma  \gamma'$ conversion after propagating through distance $z$ in a vacuum is given by
\begin{equation}
P_{\gamma \gamma'}(z)=4\chi^2 \sin^2 \left( \frac{m^2_{\gamma'}}{4 \omega } z \right),
\end{equation}
where $\omega$, $\chi$, and $m_{\gamma'}$ are the photon energy, coupling constant, and a mass of the hidden-sector photon, respectively. 
The hidden-sector photon reconverts into a photon after propagating through a wall. 
This scheme is known as the ``light shining through a wall (LSW)'' experiment~\cite{Sikivie:2007qm, Okun:1982xi}.
The probability of the photon existing through a wall is given by
\begin{eqnarray}
P_{trans}&=&P_{\gamma \gamma'}(L_1)P_{\gamma' \gamma}(L_2)  \nonumber \\
&=&16 \chi^4  \sin^2 \left( \frac{m^2_{\gamma'}}{4 \omega}L_1 \right)\sin^2 \left( \frac{m^2_{\gamma'}}{4 \omega}L_2 \right), \label{trans}
\end{eqnarray}
where $L_1$ and $L_2$ are the lengths of the conversion and re-conversion regions, respectively. 
Many LSW experiments have been proposed and implemented on a microwave light and laser at $\lambda = 1064~\mathrm{nm}$~\cite{Bahre:2013ywa, Betz:2013dza}. 
The ALPS experiment obtained a coupling constant of $\chi <1.4 \times 10^{-6} $ at a mass of $1.8 \times 10^{-4}$~eV within $95\%$ confidence level limits~\cite{Ehret:2010mh}. 
However, the ALPS sensitivity in the low-mass region is limited by the lengths of the conversion and re-conversion regions. 
Currently, the ALPS experiment is being upgraded to ALPS-IIb~\cite{Bahre:2013ywa}.\\
This paper investigates whether a long-baseline cavity with a high-powered laser in a laser interferometric gravitational-wave (GW) detector is suitable for the hidden-sector photon experiments.
The GW sensitivity is conferred by two state-of-the-art long-baseline cavities and a high-powered laser. The laser generates a large number of the hidden-sector photons.
Therefore, the combination of laser-cavity and high power laser is an excellent source of the hidden-sector photons. When a reconversion region with a reconversion cavity (RC) is placed at the backside of an arm in the GW detector, these hidden-sector photons might be reconverted to the visible photons. Figure~\ref{fig:LSW} illustrates concept of our proposed experiment based on KAGRA~\cite{Somiya:2011np,Aso:2013eba}.  In this paper, we present the expected sensitivity of the hidden-sector photon searching, using several types of the long-baseline cavity.\\

\noindent
\section{Assumption} The sensitivity to the coupling constant and the mass of the hidden-sector photon is evaluated in three GW experiments; Advanced LIGO (AdvLIGO)~\cite{TheLIGOScientific:2014jea}, KAGRA~\cite{Somiya:2011np,Aso:2013eba}, and 
 Einstein Telescope (ET)~\cite{Hild:2010id}.
 AdvLIGO is the second-generation GW experiment realized by two interferometers. These are located in Hanford, WA and Livingston, LA, USA, each with 4~km Fabry-Perot cavities. 
The optical source of both interferometers is a 532 and 1064~nm Nd:YAG laser. 
Using a power-recycling scheme~\cite{TheLIGOScientific:2014jea}, the effective input power is approximately 5.2~kW per cavity. 
The Fabry-Perot cavity stores the laser power using highly reflective mirrors. 
The average number of the laser reflections is 287, providing a stored laser power of 745~kW. 
The second-point-five generation GW experiment, KAGRA, employs the mirrors at the 20~K to reduce thermal noise. 
The cavity length in the experiment is 3~km and the total stored laser power is 410~kW. 
In ET proposed as the third-generation GW experiment, the telescope detector is split into two interferometers optimized to low-frequency and high-frequency GWs, called ET-D-LF and ET-D-HF, respectively.  The cavity length of 10~km and the stored power of 10~MW with the ET-D-HF are ideally suited for studying the hidden-sector photons. The parameters of the GW experiments are listed in Table~\ref{tab:GW}. 

\begin{table}
\begin{center}
 \caption{Specifications of ALPS-IIb experiments and their long-baseline cavities. Here, power, \#reflection, and base-line are initial power input to the cavity, number of reelections in the cavity, and denotes length of cavity, respectively.   \label{tab:GW}	}
\begin{tabular}{c|c|c|c|c} 
Project & ALPS-IIb & AdvLIGO& KAGRA& ET\\ \hline
\footnotesize{Wavelength} &$1064\mathrm{nm}$&$1064~\mathrm{nm}$&$1064~\mathrm{nm}$&$1064~\mathrm{nm}$ \\
Power &30~W&5.2~kW&825~W&500~W \\
Finesse &7859& 450 &1550&- \\
\#reflection &5000&287&987&-\\
Stored power &150~kW&745~kW&410~kW&10~MW \\
baseline &100~m&4000~m&3000~m&10~km
 \end{tabular}
 \end{center}
\end{table}

The transition edge sensor (TES) bolometer~\cite{Bahre:2013ywa} is used as a reconverted photon detector
at the end of the reconversion region. When the additional optical power is deposited on the TES detector, the bolometer island heats up, increasing the resistance.  
We can read a power during the steep resistance change.  
The TES bolometer detects the small temperature changes as the photons are absorbed and converted to heat. 
In a realistic configuration, we use fibers as the photons are translated to the TES bolometer.
the one-photon detection efficiency with fibers is $\eta=0.75$~\cite{Bahre:2013ywa}. 
The detected number of the regenerated photons on the TES detector is given by
\begin{equation}
S_{tot}(\eta,m_{\gamma'})=\eta N_0 \left[ \frac{(N_{1}^{pass}+1)(N_{2}^{pass}+1)}{4} \right] P_{trans},  \label{tot}
\end{equation}
where $N_0$ is the number of the photons in the incident beam, and $N_{1}^{pass}$ and $N_{2}^{pass}$ are the numbers of laser reflections in the RC and the conversion cavity (CC), respectively~\cite{Sikivie:2007qm}. 
The incident beam is obtained by $N_0=P_0/w$, where $P_0$ is an initial power input to the cavity.
The stored laser power is defined by $P_{arm}=P_0 \times N_1$.
Each the number of refraction correspond to finesse $F_i$ as $N_i^{pass}=2F_i/\pi$ $(i=1,2)$. 
When we these cavities are unused, we have $N_{1}^{pass}=0$ and $N_{2}^{pass}=0$.
The dark rate of TES is $B=0.000001~\mathrm{s^{-1}}$~\cite{Bahre:2013ywa}. \\

\begin{figure}
\begin{center}
\includegraphics[width=12cm]{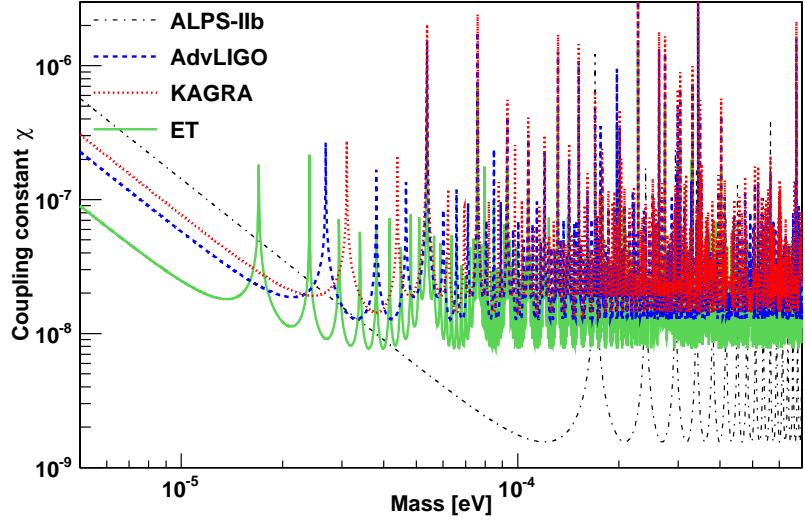}
\caption{Projected sensitivity of the LSW experiments to $\gamma \gamma'$ oscillations detected by the TES bolometer and the converted region of 1000~m. The y-and x-axes represent the coupling constant and mass of hidden-sector photon, respectively. The measurement time is assumed as 1 year.  Black dash-dotted curve is the expected sensitivity of the future ALPS-IIb experiment with a 100~m cavity. Dashed blue, dotted red, and solid green curves are the sensitivities of AdvLIGO, KAGRA, and ET, respectively. \label{fig:sensitivity1}}
\end{center}
\end{figure}

\begin{figure}
\begin{center}
\includegraphics[width=12cm]{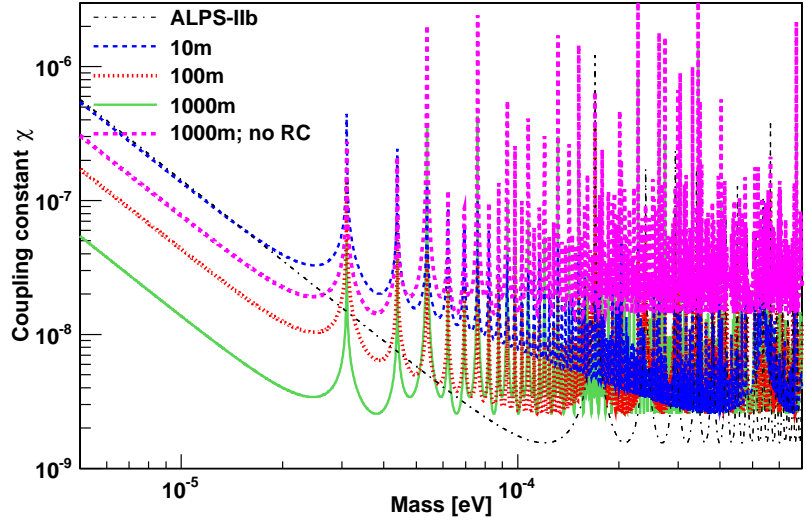}
\caption{ Same as Figure~\ref{fig:sensitivity1} except the sensitivity of KAGRA with and without the RC. Black dash-dotted curve is the expected sensitivity of the future of ALPS-IIb experiment with a 100~m cavity. Dashed blue, dotted red, and solid green curves are sensitivities when the length of the reconverted region in the RC is 10~m, 100~m and 1000~m, respectively. The dashed purple curve is the sensitivity without the RC. \label{fig:sensitivity2} }
\end{center}
\end{figure} 
\begin{figure}
\begin{center}
\includegraphics[width=12cm]{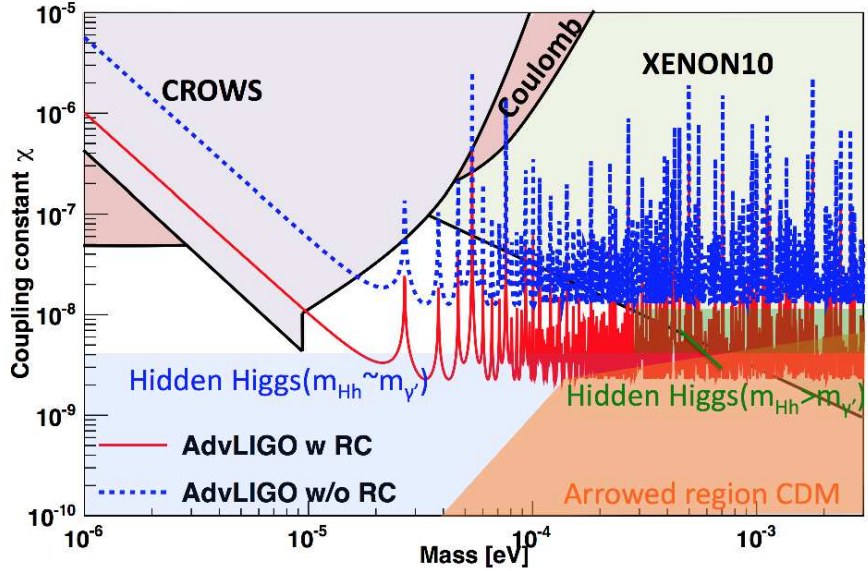}
\caption{Same as Figure~\ref{fig:sensitivity1} except the previous experiment and the allowed regions of the hidden Higgs and cold dark matter~\cite{Arias:2012az,Goodsell:2009xc}. The AdvLIGO sensitivity is over-plotted in same figure. theXENON10 is a new dark matter search experiment, aiming to increase the fiducially liquid xenon~\cite{An:2013yua}. The  CROWS is microwave based LSW experiments~\cite{Betz:2013dza}. The Coulomb region is excluded from tests of the inverse square law of the Coulomb interaction~\cite{PhysRevLett.26.721}.
\label{fig:sensitivity3} }
\end{center}
\end{figure}

\noindent
\section{Sensitivity and discussion} To assess the expected sensitivity, we assume that 
the number of the detected photons is the sum of the expected signal $S_{tot}\times T$ and the background $B \times T$, where $T$ is the observation time. 
From the background, we estimate the number of the detected photons to be $2\sigma$ significance level.
The sensitivity curve as a function of $\chi$ is calculated with $2\sigma$ significance by the term of $(S_{tot}+B)T$.  
The expected sensitivities of ALPS-IIb, AdvLIGO, KAGRA, and ET are plotted in Figure~\ref{fig:sensitivity1}.
The length of each the reconversion regions is assumed as 1000~m.
We find that AdvLIGO is slightly more sensitive than KAGRA.
This is because that the stored power of AdvLIGO is larger than that of KAGRA.
The most sensitive experiment is ET, by the virtue of its power and long-baseline cavity.
Moreover, these sensitivities are expected to exceed that of the ALPS-IIb experiment, with $m_{\gamma'}$ of the order of $10^{-5}$.

We evaluated the effect of the RC on the sensitivity although the installation is very challenging.
This is because that we have to align the large mirrors while tuning the phase and the beam shape to resonance. Furthermore, we have to achieve 1000 reflection. We discuss the possibility of RC in the case of KAGRA.
In order to realize RC, we should resolve three points:
\begin{itemize}
\item the alignment,
\item the large mirror and the anti-reflection (AR) coating,
\item the resonant condition.
\end{itemize}
First, we consider to the possibility of the alignment.
The KAGRA places the beam reducing telescope (BRT) behind the end-mirror in the arm.
When we align the mirror of RC, we use the split beam from the BRT as a reference.
This is because that this beam aligns the optical axis of CC. 
In this way, we realize the alignment of the long baseline resonator.
Second, we examine the possibility of the large mirror and the anti-reflection (AR) coating.
The beam spot diameter of KAGRA is less than 60~mm with 1~km points from the start of the reconversion region. 
Therefore, we need the large mirror with a diameter of 300 mm, which correspond to the beam width with $5 \sigma$.  
The LIGO experiment employ the AR-coated fused silica mirrors with a diameter of  340~mm~\cite{0264-9381-27-8-084006}.
Therefore, we can make the sufficient large mirrors.
Third, we should discuss the resonant condition.
We can apply the green lock laser, placed in the ALPS-II experiment~\cite{Bahre:2013ywa}.
The beam in BRT can divide the beam splitter.
The split beam is converted the green laser while keeping the phase of beam.
The green laser is injected to RC and locking to resonant condition.
Therefore, the technology of RC has a sufficient realization.

In the study,  the refraction time of RC was assumed as 1000.
Adopting this cavity, the sensitivity of KAGRA dramatically improved by a factor of six as shown in Fig.~\ref{fig:sensitivity2}. 
From Eq. (\ref{tot}), the number of reflections is proportional to $\chi^{4}$.
The RC length from 10~m to 1000~m improves the sensitivity by  three orders of magnitude.
On the other hand, the conversion and reconversion length depend on the mass scale.
The mass cut-off at these length is proportional to $1/L_1$ and $1/L_2$, respectively.
However, in the real case, the mass cut-off is derived by combining the length of both regions.

Figure~\ref{fig:sensitivity3} shows the dark matter predictions and the previous experiments.
The AdvLIGO (reconversion region=1000 m) is over-plotted in same figure. 
As shown in Fig.~\ref{fig:sensitivity3}, the AdvLIGO results are expected to surpass the previous exclusion limits . 
In the second-point-five generation GW experiments, AdvLIGO is expected to achieve a coupling constant $\chi = 1 \times 10^{-8}$ at a mass of $ 2 \times 10^{-5}$~eV, even without the RC.
By placing the RC in the conversion region, this result should improve to $\chi = 2.0 \times 10^{-9}$ , implying that we can improve the XENON10 limits by over one order of magnitude.
Furthermore, we can close the gap between the XENON10 and CROWS limits.
These results may be constrained by string physics and cold dark matter.
We superimpose the arrowed regions of these predictions in Fig.~\ref{fig:sensitivity3}.
The blue and green regions are predicted by the beyond the standard-model of the string theory~\cite{Goodsell:2009xc}.
The blue areas correspond to the hidden Higgs mass which is
similar to a hidden photon mass $(m_{H_h} \sim m_{\gamma'})$. The green areas correspond to models with a chiral
Higgs particle, in which a mass hierarchy exists $(m_{H_h} > m_{\gamma'})$.
The orange region delineates the allowed region of the cold dark matter.
This region is constrained by the effective number of neutrinos, which measured by WMAP which constitute the cosmic microwave background~\cite{0067-0049-192-2-18}.
The motivations for these scientific exploits are detailed elsewhere~\cite{Arias:2012az,Goodsell:2009xc}.\\

\noindent
\section{Conclusion} We suggest a new application for interferometers used in the GW experiments.
The TES bolometer is sufficiently sensitive to search for the low-mass hidden-sector photons, which are candidates of the hidden Higgs and/or cold dark matter.
Even if we cannot measure significant signals, the sensitivity calculations suggest that we can expand the parameter space of the current measurements.
  \\

\noindent
\section{Acknowledgments}
We are also thankful to Koji Arai and for this expertise with Advanced LIGO.
We are thankful to Yuta Michimura and for this expertise with KAGRA.
We are thankful to Aine Kobayashi for the comment to study of science motivation.
The authors would like to thank Toshikazu Suzuki and Takayuki Tomaru for commenting on this study.
The suggestions of Akiteru Takamori were very helpful in clarifying the paper.
We thank Masaya Hasegawa and Shugo Oguri for the critical reading of the manuscript.
We extend my appreciation to Masashi Hazumi for their assistance and support.
The authors were supported by JSPS KAKENHI Grant Numbers $25\cdot3626$ and 25707014.


\begin{thebibliography}{99}
\bibitem{Ahlers:2007rd}  M. Ahlers, H. Gies, J. Jaeckel, J. Redondo, and A. Ringwald. Phys. Rev. D {\bf 76} 115005 (2007).  
\bibitem{PhysRevLett.38.1440}  R. D. Peccei and Helen R. Quinn. Phys. Rev. Lett., {\bf 38}, 1440-1443, (1977).
\bibitem{PhysRevLett.40.223} S. Weinberg. Phys. Rev. Lett. {\bf 40}. 223-226, (1978).
\bibitem{Kim19871} J. E. Kim. Physics Reports, {\bf 150}1 - 177, (1987)
\bibitem{PhysRevLett.40.279} F. Wilczek. Phys. Rev. Lett., {\bf 40}, 279-282, (1978).
\bibitem{Dienes:1996zr} Keith R. Dienes, Christopher F. Kolda, and John March-Russell. Nucl. Phys., {\bf B492} 104-118, (1997).
\bibitem{Sikivie:2007qm} P. Sikivie, D. B. Tanner, and K. Bibber.  Phys. Rev. Lett., {\bf 98} 172002, (2007).
\bibitem{Okun:1982xi} L. B. Okun. Sov. Phys. JETP, {\bf 56}, 502, (1982).
\bibitem{Bahre:2013ywa}R. Bahre et al. JINST, {\bf 8} T09001, (2013).
\bibitem{Betz:2013dza}M. Betz, F. Caspers, M. Gasior, M. Thumm, and S. W. Rieger. Phys. Rev.D. {\bf 88} 075014, (2013).
\bibitem{Ehret:2010mh} K. Ehret et al. Phys. Lett., B {\bf 689} 149-155, (2010).
\bibitem{Somiya:2011np} K. Somiya. Class. Quant. Grav., {\bf 29} 124007, (2012).
\bibitem{Aso:2013eba} Y. Aso et al.,Phys. Rev. D {\bf 88(4)}, 043007, (2013).
\bibitem{TheLIGOScientific:2014jea} J. Aasi et al., Class. Quant. Grav., {\bf 32} 074001, (2015).
\bibitem{Hild:2010id} S. Hild et al. Class. Quant. Grav., {\bf 28} 094013, (2011).
\bibitem{Arias:2012az} P. Arias, D. Cadamuro, M. Goodsell, J. Jaeckel, J. Redondo, and A. Ringwald. JCAP, {\bf 1206} 013, (2012).
\bibitem{0264-9381-27-8-084006} Gregory M Harry (forthe LIGO Scientific Collaboration).Classical and Quantum Gravity, {\bf 27-8}, 084006,(2010).
\bibitem{Goodsell:2009xc} M. Goodsell, J. Jaeckel, J. Redondo, and A. Ringwald. JHEP, {\bf 11} 027, (2009).
\bibitem{An:2013yua} H. An, M. Pospelov, and J. Pradler. Phys. Rev. Lett., {\bf 111} 041302, (2013).
\bibitem{PhysRevLett.26.721} E. R. Williams, J. E. Faller, and H. A. Hill. Phys. Rev. Lett. {\bf 26} 721-724, (1971).
\bibitem{0067-0049-192-2-18} E. Komatsu et al.,The Astrophysical Journal Supplement Series, {\bf 192(2)} 18, (2011).

\end{thebibliography}
\end{document}